\documentclass[twocolumn, aps, pra, superscriptaddress,reprint]{revtex4-2}

\usepackage{graphicx}
\usepackage{multirow}
\usepackage{amsmath,amssymb,amsfonts}
\usepackage{amsthm}
\usepackage{mathrsfs}
\usepackage[title]{appendix}
\usepackage{xcolor}
\usepackage{textcomp}
\usepackage{booktabs}
\usepackage{algorithm}
\usepackage{algorithmicx}
\usepackage{algpseudocode}
\usepackage{listings}
\usepackage{siunitx}

\theoremstyle{plain}

\theoremstyle{definition}

\theoremstyle{remark}

\begin{document}

\title[Article Title]{Multi-kilohertz laser plasma acceleration driven by an industrial-grade Yb:YAG laser}

\author{B. Farace}
\email{bonaventura.farace@desy.de}
\affiliation{Deutsches Elektronen-Synchrotron DESY, Notkestrasse 85, 22607 Hamburg, Germany}

\author{N. Khodakovskiy}
\affiliation{Deutsches Elektronen-Synchrotron DESY, Notkestrasse 85, 22607 Hamburg, Germany}

\author{R.J. Shalloo}
\affiliation{Deutsches Elektronen-Synchrotron DESY, Notkestrasse 85, 22607 Hamburg, Germany}

\author{T.G. Pak}
\affiliation{Deutsches Elektronen-Synchrotron DESY, Notkestrasse 85, 22607 Hamburg, Germany}

\author{E. Escoto}
\affiliation{Deutsches Elektronen-Synchrotron DESY, Notkestrasse 85, 22607 Hamburg, Germany}

\author{S. Rajhans}
\affiliation{Deutsches Elektronen-Synchrotron DESY, Notkestrasse 85, 22607 Hamburg, Germany}

\author{A. Sch\"{o}nberg}
\affiliation{Deutsches Elektronen-Synchrotron DESY, Notkestrasse 85, 22607 Hamburg, Germany}

\author{I. Hartl}
\affiliation{Deutsches Elektronen-Synchrotron DESY, Notkestrasse 85, 22607 Hamburg, Germany}

\author{J. Osterhoff}
\affiliation{Deutsches Elektronen-Synchrotron DESY, Notkestrasse 85, 22607 Hamburg, Germany}

\author{C.M. Heyl}
\affiliation{Deutsches Elektronen-Synchrotron DESY, Notkestrasse 85, 22607 Hamburg, Germany}
\affiliation{GSI Helmholtzzentrum für Schwerionenforschung GmbH, Planckstraße 1, 64291 Darmstadt, Germany}
\affiliation{Helmholtz-Institute Jena, Fröbelstieg 3, 07743 Jena, Germany}

\author{A.R. Maier}
\affiliation{Deutsches Elektronen-Synchrotron DESY, Notkestrasse 85, 22607 Hamburg, Germany}

\author{K. P\~{o}der}
\affiliation{Deutsches Elektronen-Synchrotron DESY, Notkestrasse 85, 22607 Hamburg, Germany}

\author{W. P. Leemans}
\affiliation{Deutsches Elektronen-Synchrotron DESY, Notkestrasse 85, 22607 Hamburg, Germany}
\affiliation{Department of Physics, University of Hamburg, Mittelweg 177, 20148 Hamburg, Germany}


\begin{abstract}
{Laser plasma accelerators (LPAs) are a promising platform for compact radiation sources. For a wide range of applications, including radiotherapy, ultrafast electron diffraction and time-resolved imaging, stable operation at high repetition rates is essential in order to deliver competitive average particle flux. Here we demonstrate the first LPA driven by an industrial-grade ytterbium-doped yttrium aluminium garnet (Yb:YAG) laser, designed for high-average-power operation. The picosecond laser pulses are post-compressed in a multi-pass cell to 50 fs duration and used to drive the interaction. The electron accelerator is operated in burst mode, at repetition rates tuneable from 0.625 to 6.25 kHz, representing a substantial increase compared to the state-of-the-art. Across this range, the electron beam properties remain unchanged, with average charges of 10-12 pC per shot, divergences of 50-70 mrad, and Maxwellian-like spectra extending to a few MeV. Numerical simulations capture the key features of the experimental observations and indicate acceleration in the self-modulated regime, enabled by relativistic self-focusing in near-critical-density plasma. Combining industrial high-average-power laser technology with plasma-based acceleration, these results represent a key step toward scalable, compact high-repetition-rate electron sources for medical, imaging and industrial applications.}
\end{abstract}

\keywords{Laser plasma accelerators, High-repetition-rate, Yb:YAG, Compact electron source}



\maketitle

\section{Introduction}\label{sec:intro}

Achieving robust, scalable, high-repetition-rate operation remains a central challenge in transitioning laser plasma accelerators from laboratory setups to practical applications. A laser plasma accelerator (LPA) exploits the interaction of a laser pulse of relativistic intensity with a tenuous plasma to excite strong accelerating gradients, reaching hundreds of gigavolts per metre \cite{Tajima1979, Leemans2006, Leemans2014, Poder2024}. This, combined with the possibility of producing ultrashort electron bunches of femtosecond duration, has established over the last decades LPAs as a promising technology for next-generation electron accelerators. Their characteristics make them particularly attractive candidates in a broad range of applications, including ultrafast electron diffraction, time-resolved imaging and radiotherapy \cite{Glinec2005, He2013a, He2016, Andreassi2016, Hidding2017, Cavallone2021}, provided that sufficiently high repetition rates and average beam current can be achieved. To date, most LPAs rely on 100 TW class Titanium:Sapphire lasers owing to their broad gain bandwidth, which supports short pulse durations ($\sim$\,30 fs) and enables high output peak power. While these systems typically operate at $\sim$\,1 Hz, important progress towards higher repetition rates has already been achieved: 1 kHz-level LPA operation has been demonstrated in several experiments, establishing the feasibility of laser-driven electron acceleration in this regime \cite{Beaurepaire2014, Faure2018, Guenot2017, Gustas2018, He2013b, Rovige2020, Rovige2021, Salehi2017, Salehi2021}. However, despite these advances, further scaling remains challenging due to demanding thermal management requirements, arising from the pumping architecture and the large quantum defect of Ti:Sa crystals \cite{Wolter2017}. This poses severe constraints on the achievable average power, ultimately limiting their practical use. As a consequence, applications requiring sustained, scalable, high-throughput operation motivate the search for a more efficient and robust laser technology.
In this context, Ytterbium technology is particularly well-positioned: Yb-doped materials feature a lower quantum defect, a comparably long upper-state lifetime and can be directly pumped by high power, high efficiency diodes, enabling efficient high-average-power operation \cite{Wolter2017, Rothhardt2017}. In particular, Yb:YAG (yttrium aluminium garnet) lasers can deliver millijoule-level pulses at multi-kilohertz repetition rates with long-term stability and industrial-grade robustness \cite{Russbueldt2009}. Therefore, these systems represent a promising, novel alternative as drivers for high-average-power, few MeV LPAs. Nevertheless, their limited intrinsic gain bandwidth and consequently long output pulse duration ($\sim$\,1 ps) have so far hindered their use. As a result, developing LPA-technology leveraging industrial Yb-based lasers has remained largely unexplored. Recent advances in, multi-pass cell (MPC) post-compression schemes have, however, enabled efficient pulse compression to the tens of femtoseconds regime, thereby bridging this gap \cite{Hanna2021, Kaumanns2021, Viotti2022, Pfaff2023, Rajhans2023}.
In the following, the first laser-plasma accelerator driven by an industrial-grade Yb:YAG laser, post-compressed in a MPC-based setup is presented.
Electron acceleration is demonstrated in burst-mode with stable beam characteristics at repetition rates up to 6.25 kHz, exceeding the current state-of-the-art by more than a factor six. Across the explored repetition rate range, the beam delivers 10-12 pC per shot, with a divergence of 50-70 mrad (half-angle) and a Maxwellian spectral distribution extending to 4 MeV. These results unequivocally demonstrate the potential of industrial high-average-power laser technology as a viable driver for plasma-based accelerators, providing a scalable route toward application-oriented LPAs.

\begin{figure*}[ht]
    \centering
    \includegraphics[width=1\linewidth]{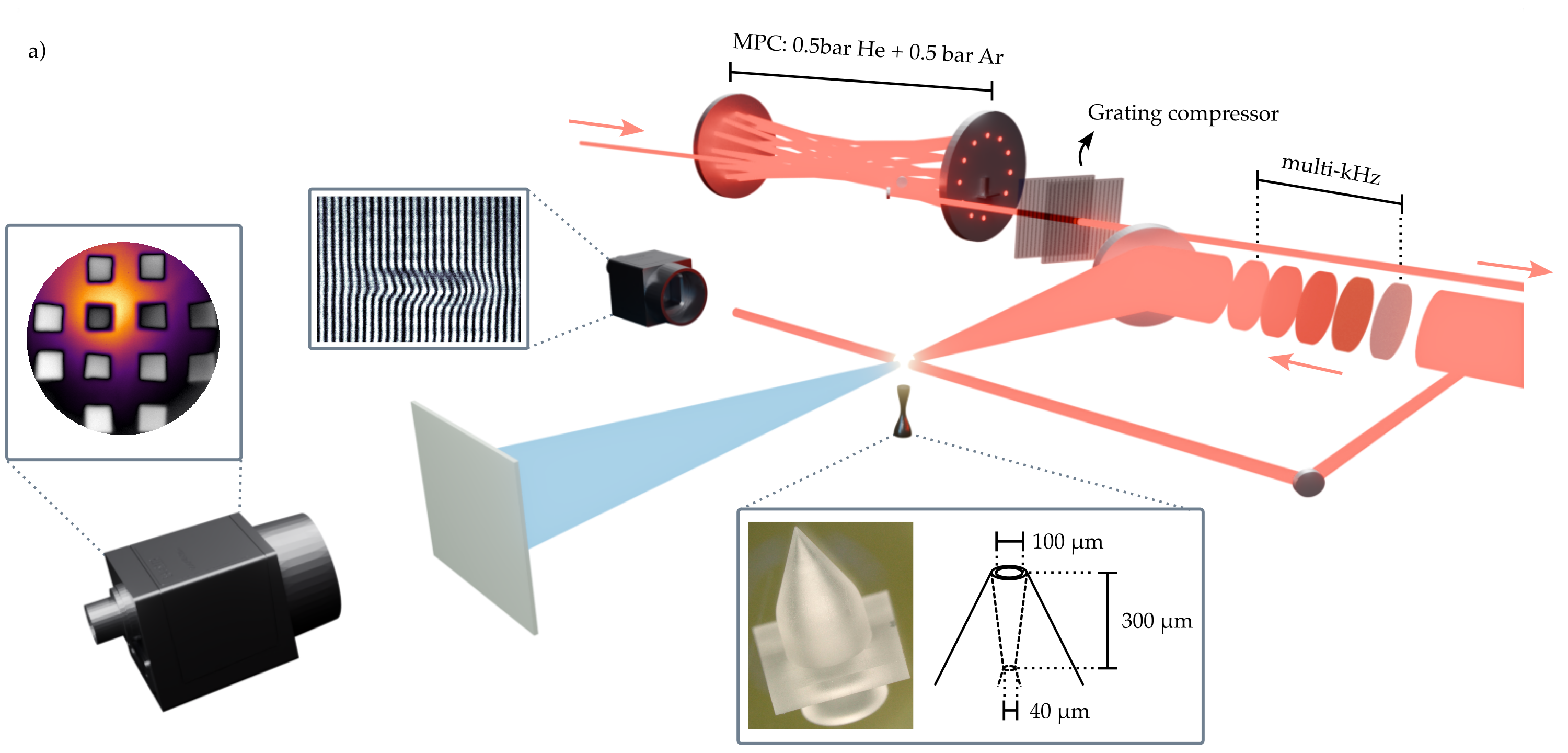} \\
    \includegraphics[width=1\linewidth]{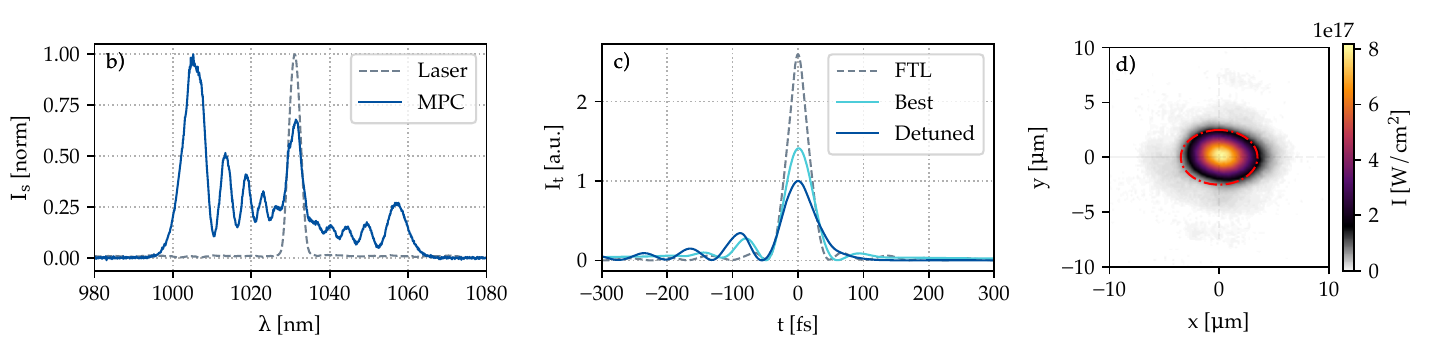}
    \caption{Panel a: Experimental setup as described in the main text. The insets show, from left to right: the typical scintillator screen image, a sample interferometry image and the supersonic nozzle used during the experiment. Panel b: laser spectrum at the output of the laser (grey) compared to the broadened spectrum after the multi-pass cell (blue). Panel c: temporal intensity of the pulse at the interarction point. The fourier transform limit (FTL, $\sim$\,41 fs FWHM), the best compression ($\sim$\,50 fs) and the detuned compression optimised for electron acceleration ($\sim$\,70 fs) are shown in grey, light blue and blue, respectively. Panel d: measured focal spot at the interaction point. \label{fig:setup_and_laser}}
\end{figure*}

\section{Results}\label{sec:results}

\subsection{Experimental setup}\label{subsec:setup}

A sketch of the experimental setup is presented in Fig.~\ref{fig:setup_and_laser}a. An industrial-grade Yb:YAG laser (Amphos 3000), with central wavelength $\lambda_0 = 1030$ nm, is used as driver for the laser plasma interaction. Nominally, the laser delivers pulses of 16.5 mJ and 1.4 ps FWHM at 12.5 kHz, corresponding to a maximum average power exceeding 200 W. The repetition rate can be adjusted via integer division of the 12.5 kHz base frequency, allowing operation down to 0.625 kHz. In this work, the LPA performance is investigated at repetition rates of 0.625, 1.25, 4.167, and 6.25 kHz, spanning one order of magnitude. For clarity, these values are hereafter referred to as 0.6, 1, 4, and 6 kHz.

At the output of the laser, a MPC-based post-compression setup, adapted from the scheme detailed in \cite{Rajhans2023}, is used to achieve the relativistic intensity required for the laser plasma interaction \cite{Viotti2022}. The broadening of the initially narrow spectral bandwidth, represented in Fig.~\ref{fig:setup_and_laser}b, is achieved through self-phase modulation in a 2 m-long cell filled with a mixture of argon and helium gases at atmospheric pressure. Afterwards, the broadened pulse is transported through a $\sim$ 35 m beamline and compressed by a single-pass transmission grating compressor (Coherent LightSmyth), yielding a final pulse duration of about 50 fs, as shown in Fig.~\ref{fig:setup_and_laser}c. Although a single-pass configuration inherently introduces spatial chirp, this effect remains negligible owing to a large beam radius (11.5 mm $\times$ 10 mm 1/e$^2$) and to a small grating separation ($\sim$ 2 mm). Moreover, the single-pass geometry provides a compact and easily adjustable setup with high transmission efficiency (96\%), offering an effective alternative to chirped-mirror-based compression schemes.

Following post-compression, the laser is sent to the interaction chamber and focused onto the plasma target using an f/2.5 off-axis parabolic mirror (f = 50.8 mm, Thorlabs), mounted on a miniature hexapod stage (H-811, PI). Fine optimisation of the focal spot is achieved via a deformable mirror (Ilao-Star 50, Imagine Optics), resulting in a typical spot size of \SI{3.5}{\micro\metre} $\times$ \SI{2.5}{\micro\metre} 1/e$^2$ radius (in vacuum). Fig.~\ref{fig:setup_and_laser}d shows the laser intensity at the interaction point reaching $\sim$~8 × 10$^{17}$ W cm$^{-2}$, corresponding to a normalised (vacuum) vector potential $a_0 = p/m_ec\sim0.8$, where $m_e$ is the electron rest mass, $p$ is its momentum and $c$ is the speed of light in vacuum. The parameter $a_0$ characterises the maximum momentum acquired by an electron oscillating in the laser field, normalised to $m_ec$, and provides a measure of the relativistic strength of the interaction. Here, where $a_0<1$, the interaction would remain only mildly relativistic, and electron injection into the plasma wakefield would not occur without additional enhancement mechanisms.
Therefore, to achieve injection despite the modest laser intensity, the plasma target is designed to operate at electron plasma densities approaching the critical density $n_c\sim1\times10^{21}$ cm$^{-3}$, defined as the density above which electromagnetic radiation can no longer propagate in the medium. In this density regime relativistic self-focusing can strongly enhance the laser field during propagation, resulting in a significant increase in its $a_0$ \cite{Sun1987}. To this end, a supersonic, fused-silica, de Laval nozzle \cite{Schmid2012} has been designed in-house to produce the required density profile and manufactured by LightFab Gmbh. The nozzle geometry, shown in the inset of Fig.~\ref{fig:setup_and_laser}a, is tailored to the experimentally measured laser parameters. Its design is a result of a Bayesian optimisation loop combining computational fluid dynamics (CFD) and particle-in-cell (PIC) simulations, aimed at maximising the injected charge, as detailed in \cite{Farace2024}. By adopting this plasma source and using pure nitrogen as gas, relativistic self-focusing increases the effective normalised vector potential during propagation and enables electron injection in the self-modulated wakefield regime \cite{Esarey2009}, as discussed in Section \ref{sec:discussion}. 

The gas delivery system is designed to sustain stable backing pressure at the nozzle inlet up to 45 bar, corresponding to a maximum peak plasma electron density of approximately 3.5 $\times$ 10$^{20}$ cm$^{-3}$ along the laser axis. The resulting substantial gas load prevents continuous gas operation during the experiment. Therefore, during the data acquisition, the source is operated in a pulsed mode, at 0.7 Hz and with an opening time of 15 ms. Nevertheless, within each gas burst, the plasma is continuously replenished and electron acceleration takes place on every laser shot, resulting in effective multi-kHz electron acceleration. The overall duty cycle, currently limited by the vacuum system capacity, is expected to be substantially increased with a differential pumping system, eventually enabling continuous, high-density operation \cite{Monzac2025}.

Before the focusing parabola, a small fraction of the main laser is picked off and used as a transverse probe in a Mach–Zehnder interferometer, allowing the reconstruction of the plasma density profile (cf. Methods). Following the interaction, the laser rapidly diffracts and is subsequently blocked by a thin aluminium foil. On the other hand, the accelerated electron beam propagates towards a 50 mm diameter DRZ-High scintillating screen, positioned 16.5 cm downstream, for characterisation. The screen, calibrated with a tritium source of known activity, is equipped with twelve 5 mm $\times$ 5 mm tungsten filters of varying thicknesses (from 25 to 300 $\mu$m) for the reconstruction of the electron energy spectrum (cf. Methods).

\subsection{Electron acceleration}\label{subsec:electron_acceleration}

\begin{figure}
    \centering
    \includegraphics[width=1\linewidth]{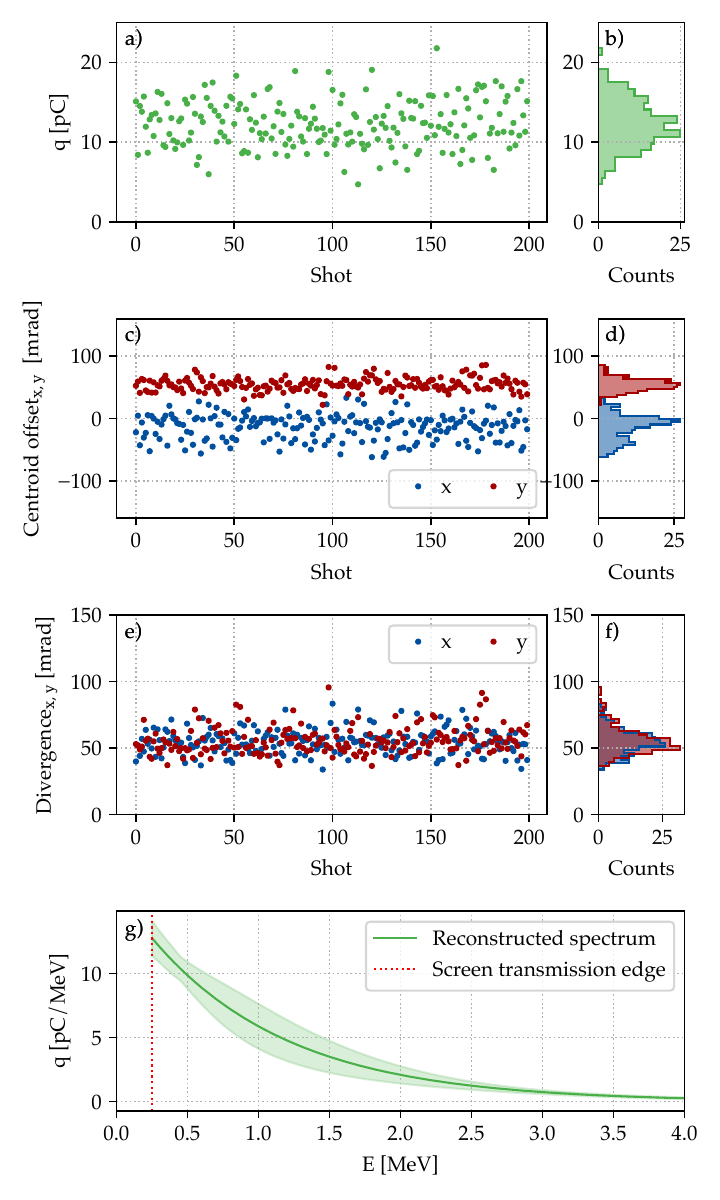}
    \caption{Electron beam charge, pointing, divergence and reconstructed spectrum for a nozzle backing pressure of 45 bar and with the laser operating at 4 kHz. Panels a-b: accelerated charge for 200 laser shots with the nozzle placed at the optimum position, each acquired from an individual gas burst. Panels c-f: shot-to-shot pointing and divergence on x (red) and y (blue) for the same 200 shots. Panel g: reconstructed electron spectrum, averaged over the 200 shots. The red dotted line shows the minimum detectable energy.}
    \label{fig:4kHz_signal}
\end{figure}

The key advantage of multi-kHz operation lies in increasing the average electron flux. Accordingly, the experimental optimisation focuses on maximising the accelerated charge per laser pulse, while maintaining a robust shot-to-shot reproducibility. 

To this aim, achieving the highest possible peak intensity during the interaction is crucial, as both the injected charge and the amplitude of the plasma wakefield strongly depend on it. At the near-critical plasma densities considered in this work, the refractive index of the plasma can significantly affect the spectral phase of the laser pulse. As a result, even small variations in the initial group delay dispersion (GDD) can modify the nonlinear laser-plasma coupling and, consequently, the electron injection process \cite{Shalloo2020}. Therefore, in a first step, the phase of the driving pulse is adjusted by fine-tuning the gratings separation in the compressor, thereby controlling the residual GDD at the interaction point. The compressor configuration yielding the highest and most stable accelerated charge is then selected for all the subsequent measurements. As shown in Fig.~\ref{fig:setup_and_laser}c, this operating point corresponds to a FWHM pulse duration of $\sim$\,70 fs and an estimated GDD of +110 fs$^2$ with respect to the transform limited pulse duration. Following the spectral phase optimisation, the electron signal is systematically studied as a function of the relative position between the gas nozzle and the laser focal position, with varying nozzle backing pressure and for different laser repetition rates. The diagnostics are synchronised to the main laser oscillator and gated to record a single shot at each acquisition, allowing shot-resolved measurements over the full repetition-rate range. Each acquisition is triggered by the gas nozzle, such that one laser shot is selectively recorded per gas burst. In the subsequent analysis of the measured electron data, only shots with an accelerated charge exceeding 100 fC are considered, ensuring sufficient signal for a reliable characterisation. This threshold excludes less than 0.5\% of the recorded shots. 
Fig.~\ref{fig:4kHz_signal} shows the results obtained at 4 kHz with the nozzle outlet positioned \SI{120}{\micro\metre} below the laser propagation axis. This offset corresponds to the minimum separation that ensures long-term operation without inducing severe damages to the plasma source. After the optimisation of both the longitudinal and transverse nozzle positions, consistent electron acceleration is observed starting from a minimum backing pressure of 35 bar, with an average charge per shot of 16 pC and a standard deviation of 34\%. With increasing backing pressure, the mean accelerated charge reduces, while the shot-to-shot stability improves. As shown in Fig.~\ref{fig:4kHz_signal}a-b, at the maximum available pressure of 45 bar, corresponding to an estimated peak plasma density of 2.7 $\times$ 10$^{20}$ cm$^{-3}$ $\pm$ 5$\%$, the average charge measured over 200 shots amounts to 12.3 pC, with a standard deviation of 24\%. 
Fig.~\ref{fig:4kHz_signal}c-f outlines the corresponding electron beam pointing fluctuation, defined as the offset of the beam centroid with respect to the laser axis, and its divergence. Overall, the shot-to-shot variations are comparable to those observed in the accelerated charge. The mean half-angle divergence is approximately 55 mrad $\pm$ 17\% on both axis, while the average centroid offset amounts to -14 mrad on the horizontal axis (x) and 55 mrad on the vertical one (y), with standard deviations of 21 mrad and 10 mrad respectively. The slight horizontal offset is likely due to a small residual misalignment between the laser propagation axis and the nozzle position, as supported by the distribution having a mode value close to zero. In contrast, the systematic vertical offset can mainly be attributed to the presence of a plasma density gradient along the nozzle axis. This gradient induces a corresponding variation in the plasma refractive index effectively steering the laser pulse upwards during propagation, a deflection which is subsequently imprinted on the accelerated electron beam \cite{Mittelberger2019}. 
As previously mentioned, the use of tungsten filters with varying thickness enables a quantitative reconstruction of the electron energy spectrum (cf. Methods). The reconstructed spectrum, averaged over the same 200 shots, exhibits a Maxwellian-like distribution extending to a few MeV, as shown in Fig.~\ref{fig:4kHz_signal}g. Such spectra are characteristic of multi-bucket injection in the self-modulated laser wakefield regime, as will be discussed in the next section.

\begin{figure}
    \centering
    \includegraphics[width=1\linewidth]{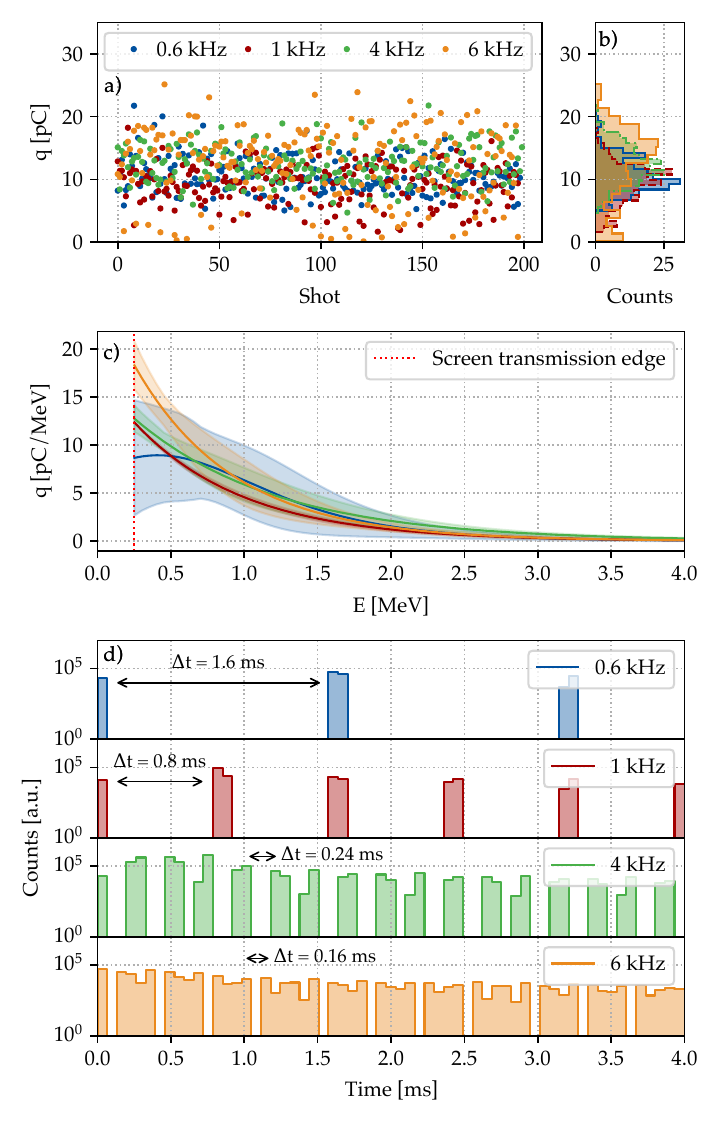}
    \caption{Electron beam characteristics across different repetition rates: 0.6 kHz (blue), 1 kHz (red), 4 kHz (green), 6 kHz (orange). Panel a-b: accelerated charge per shot for 200 shots, each acquired from an individual gas burst with a nozzle backing pressure of 45 bar. Panel c: average reconstructed electron spectrum. Panel d: time-resolved radiation detector signal, highlighting the electron acceleration frequency for the different operating regimes. At 6 kHz consecutive peaks are only partially resolved.}
    \label{fig:rep_rate_comparison}
\end{figure}

A distinctive feature of the industrial Yb:YAG laser used in this work is its broad and user-defined repetition-rate tuneability, under otherwise identical laser conditions. This characteristic provides a unique opportunity to directly assess the influence of repetition rate on the acceleration process. As shown in Fig.~\ref{fig:rep_rate_comparison}, the electron beam characteristics remain largely invariant across the explored repetition-rate range for a fixed nozzle backing pressure P = 45 bar. From 0.6 to 6 kHz no significant variation is observed in the average charge per shot, spectral distribution, or beam divergence. Specifically, the average charge per shot amounts to 10.6 pC $\pm$ 27\% at 0.6kHz, 9.3 pC $\pm$ 34\% at 1 kHz and 12.5 pC $\pm$ 45\% at 6 kHz, while the corresponding mean divergences are 64 mrad $\pm$ 22\% $\times$ 61 mrad $\pm$ 20\%, 54 mrad $\pm$ 36\% $\times$ 51 mrad $\pm$ 30\% and 65 mrad $\pm$ 29\% $\times$ 67 mrad $\pm$ 33\% respectively. The higher charge per shot fluctuations observed at 6 kHz likely originate from residual thermal effects in the multi-pass cell at the highest average powers and are currently under investigation.  

To further characterise the temporal structure of the emitted radiation, Fig.~\ref{fig:rep_rate_comparison}d outlines the time-resolved count rate recorded by a Pandora radiation detector \cite{Klett2010}. The detector is self-triggered upon radiation events and, for each trigger, acquires data over an active window of approximately 4 ms. The signal is sampled at regular intervals, corresponding to a maximum detectable frequency of $\sim$\,7.6 kHz. The Fourier analysis of the observed signal reveals the dominant radiation frequencies, which closely match the four laser operating regimes: 0.62, 1.24, 4.15 and 6.25 kHz. This provides a direct evidence that, within each gas burst, electron acceleration occurs on a shot-to-shot basis at multi-kHz frequencies. To date, 6.25 kHz represents the highest repetition rate demonstrated with laser plasma accelerators.

The demonstrated robustness of the electron beam properties across a wide multi-kilohertz regime highlights the intrinsic scalability of this approach, based on industrial high-average-power lasers and multi-pass cell post-compression. As MPC systems operating at higher average powers have already been demonstrated \cite{Mueller2021, Pfaff2023}, further optimisation of thermal management and system stability is expected to extend this performance towards the full repetition-rate capability of the laser system, paving the way to high-flux LPA-driven radiation sources.

\begin{figure}
    \centering
    \includegraphics[width=1\linewidth]{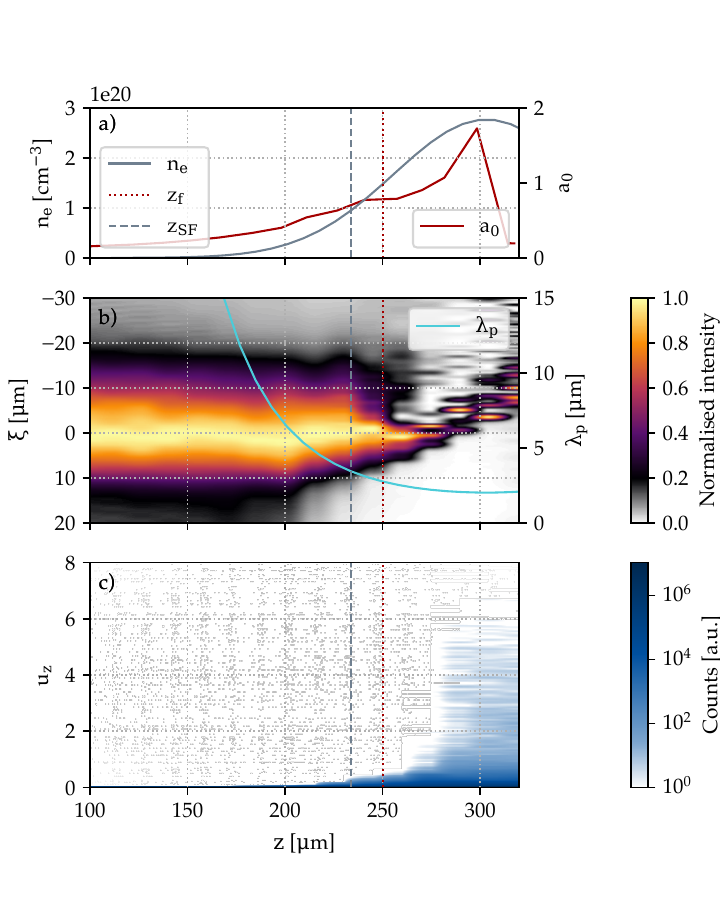}
    \caption{Laser pulse propagation in the up-ramp of the nozzle density profile simulated in FBPIC. Panel a: the peak of the normalised vector potential a$_0$ (red) is shown along the plasma density profile n$_e$ (grey) in the laboratory frame (z). The vacuum focus position of the laser z$_f$ and the theoretical position of relativistic self-focusing z$_{SF}$ are shown respectively with a red dotted and a grey dashed line. Panel b: the evolution of the laser longitudinal intensity profile during the propagation. For each simulation iteration the intensity profile is normalised to its maximum and plotted with respect to the co-moving frame $\xi = z - ct$. The plasma wavelength $\lambda_p$ is shown in light blue. Panel c: electrons normalised longitudinal momentum distribution $u_z=p_z/m_ec$. The colour map shows the number of macroparticles per momentum bin on a logarithmic scale.}
    \label{fig:fbpic_propagation}
\end{figure}

\section{Discussion}\label{sec:discussion}

To gain insight into the acceleration dynamics in this high-density, self-modulated regime and to clarify the origin of the observed spectral and stability features, a series of particle-in-cell (PIC) simulations has been performed under conditions matching the experimental parameters.

In the following, the laser plasma interaction is simulated using the LASY python library \cite{Thevenet2024} and the cylindrical quasi-3D PIC code FBPIC \cite{Lehe2016} (cf. Methods). Simulations are carried out with the measured laser focal spot in vacuum, the retrieved temporal profile of the laser pulse and the reconstructed plasma density corresponding to the optimal source position operating with a backing pressure of 45 bar. Fig~\ref{fig:fbpic_propagation} outlines the simulation results, focusing on the laser propagation during the up-ramp of the nozzle density profile. Here the vacuum focal plane is placed \SI{55}{\micro\metre} before the plasma density peak, reproducing the optimal longitudinal focus position identified experimentally. Fig~\ref{fig:fbpic_propagation}a shows the evolution of the peak of the normalised vector potential a$_0$ (red) alongside the plasma density profile (grey). Shortly before the vacuum focal position z$_f$, the combined effect of laser intensity and plasma density initiates relativistic self-focusing. For a given laser peak power, the plasma density at which self-focusing is expected can generally be estimated as

\begin{equation} \label{eq:self_focusing_density}
n_e=n_{SF}\sim 17\cdot n_c/P_p, 
\end{equation}

where $n_c \sim 10^{21}\cdot\lambda^{-2}_{L[\mu m]}$ is the critical plasma density in cm$^{-3}$ and $P_p$ is the laser peak power in GW \cite{Mori1997}. Self-focusing counteracts diffraction and further focuses the laser pulse, eventually leading to a significant increase in its a$_0$. Fig~\ref{fig:fbpic_propagation}b shows the corresponding evolution of the longitudinal laser intensity profile, normalised for each simulation step, next to the plasma wavelength $\lambda_p$ trend (light blue). As the plasma density increases, $\lambda_p$ becomes significantly shorter than the pulse length $L=c\tau_{\mathrm{fwhm}}$, where c is the speed of light in vacuum and $\tau_{\mathrm{fwhm}}$ is the temporal pulse duration. This mismatch drives two coupled mechanisms. First, in the transverse direction, the pulse propagates within the plasma wave generated by its leading edge, experiencing alternating focusing and defocusing forces arising from transverse density gradients. Second, the longitudinal plasma density modulation along the axis induces a corresponding modulation of the laser envelope. Through this coupled interaction, a positive feedback mechanism develops during the propagation, progressively increasing both the modulation of the laser envelope and the amplitude of the plasma wave. This feedback eventually leads to wavebreaking and electron injection, as illustrated in Fig~\ref{fig:fbpic_propagation}c where the evolution of the normalised longitudinal momentum distribution of the electron bunch $u_z=p_z/m_ec$ is plotted as a function of propagation distance. At z $\sim$ \SI{275}{\micro\metre} a population of electrons with $u_z>1$ abruptly appears, marking the onset of injection. As the propagation continues and the self-modulation develops further, the injected population broadens in momentum space and its mean $u_z$ increases. The resonant excitation of the wakefield typically leads to injection across multiple plasma buckets and, therefore, the distribution remains broad. In addition to this, at the near-critical plasma densities discussed in this work ($n_e/n_c \sim 0.3$)  the group velocity of the laser pulse is reduced, which significantly shortens the length over which the injected electrons are in the accelerating phase of the plasma wakefield \cite{Esarey2009}. These effects, combined with the rapid defocusing of the laser pulse following the initial strong relativistic self-focusing, limit the effective acceleration distance. As a consequence, the resulting electron spectra tend to exhibit Maxwellian-like distributions extending to few MeV, consistent with those observed experimentally and characteristic of the self-modulated acceleration regime.

\begin{figure}
    \centering
    \includegraphics[width=1\linewidth]{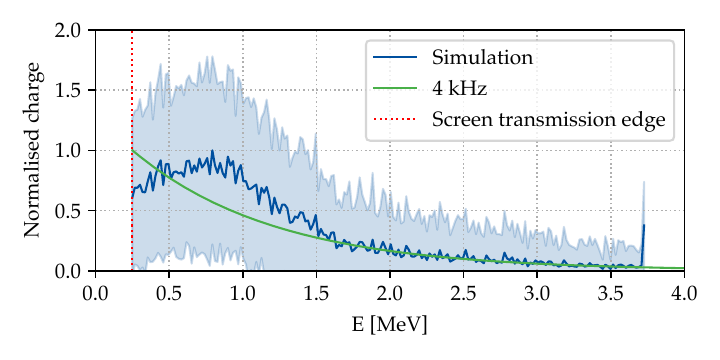}
    \caption{Normalised electron spectrum averaged over 100 FBPIC simulations at the position of the virtual scintillator screen, as described in the main text (blue). For comparison, the normalised mean reconstructed spectrum obtained at 4 kHz with 45 bar of nozzle backing pressure is plotted in green.}
    \label{fig:fbpic_grid_scan}
\end{figure}

The general agreement between simulated and measured data suggests that the model captures the essential features of the interaction. PIC simulations can therefore be used to further investigate the origin of the shot-to-shot fluctuations observed in the experiment. 
The level of fluctuations reported in the preceding section, while non-negligible, is consistent with operation in the self-modulated regime, particularly when the injection process relies on strongly nonlinear laser-plasma interaction. To identify the dominant sources of variability, a systematic analysis of the experimental parameters is performed. Typically, in LPAs, performance fluctuations are driven by instabilities of the driving laser \cite{Bohlen2022,Maier2020}. In the present case, however, the industrial-grade Yb:YAG system exhibits excellent stability. During the typical data acquisition, the relative standard deviation at the interaction point is measured to be $<$1\% in pulse energy and $\sim$ 1\% in spectral distribution, while the wavefront RMS is stable at the nm level. These values indicate that the laser driver is unlikely to be the dominant source of the observed electron-beam variations, while the major contribution likely originates from plasma density fluctuations. Operating the supersonic nozzle in pulsed, high-pressure mode, as described before, leads to $\sim$ 5\% standard deviation in the peak density and to fluctuations of $\sim$ \SI{3.5}{\micro\metre} in the nozzle position along the laser propagation axis (and thus in the relative location of the density maximum). As the interaction critically relies on relativistic self-focusing, even modest variations in plasma density or interaction length can substantially modify the nonlinear pulse evolution and lead to significant changes in the peak a$_0$ reached during the interaction. Such sensitivity naturally translates into shot-to-shot variations of the injected and accelerated charge.
To quantify this effect, a parameter scan around the experimental operating conditions is performed in FBPIC through the Optimas optimisation library \cite{FerranPousa2023}. Here, the plasma peak density and the laser focal position relative to the density maximum are varied within the measured fluctuation ranges. A total of 100 simulations are performed. Each parameter set is sampled from normal distributions centred on the mean experimental values, with standard deviations corresponding to the measured variations, thereby reproducing the shot-to-shot variability observed experimentally. 
At the end of each simulation, the accelerated electrons are ballistically propagated to a virtual detector with the same size and position as the experimental scintillator screen, enabling a direct comparison with the measurements. Fig.~\ref{fig:fbpic_grid_scan} outlines the simulations results. Owing to a large beam divergence, the electrons spatial distribution on the virtual screen appears nearly uniform, while their spectrum closely resembles the experiment. Overall, the simulations yield an average accelerated charge at the plasma exit of 26.6 pC $\pm$ 31\%. Following ballistic propagation to the virtual screen, the average detected charge decreases to 6.7 pC with a relative standard deviation of 30\%, matching the experimental variations. 
The remaining differences between simulations and experiment are likely due to the modelling assumptions. In particular, FBPIC employs a quasi-cylindrical geometry, preventing the inclusion of possible transverse asymmetries or pointing offsets between the laser pulse and the plasma density profile. In addition, the simulated plasma profile is necessarily truncated to a finite length for computational efficiency, whereas in the experiment low-density plasma tails may extend over significantly longer distances. Within these limitations, the simulations show good agreement with the experimental observations, supporting the interpretation that even modest variations in the plasma density profile can lead to the observed electron beam fluctuations. Identifying these parameters as the dominant sources of variability provides a clear path toward improved stability through enhanced plasma density control. More broadly, the robust accelerator performances demonstrated across a wide repetition-rate range, while operating with an industrial high-average-power laser, highlights the scalability of this approach and its potential for the development of high-flux, application-oriented laser plasma radiation sources.

In conclusion, this work demonstrates the first laser plasma accelerator driven by an industrial-grade Yb:YAG laser operating at multi-kHz, proving that industrial high-average-power laser technology can provide robust and scalable drivers for plasma-based radiation sources.
The original, picosecond laser pulses are post-compressed in a multi-pass cell setup to about 50 fs duration and used to drive the interaction. Electron acceleration is achieved, in burst mode, at repetition rates tuneable over one order of magnitude, from 0.625 up to 6.25 kHz, a substantial factor six increase compared with current state-of-the-art. Across this range, the beam characteristics remain stable, a key requirement for application-oriented LPA development. Few-MeV electron beams with 10-12 pC per shot and  50-70 mrad of divergence are consistently obtained in a self-modulated regime, enabled by relativistic self-focusing in near-critical plasma density.
The strongly nonlinear interaction regime, combined with burst mode gas operation, naturally leads to non-negligible shot-to-shot fluctuations, as confirmed by the combined experimental and particle-in-cell simulations analysis. These effects are associated with well-identified experimental parameters, primarily variations in the plasma density profile, and could be mitigated via technical developments. In particular, the implementation of a differential pumping scheme will enable continuous high-density gas operation, while improved thermal monitoring and optimisation of the laser compression stage are expected to mitigate residual thermal effects and extend the performances towards the full repetition-rate capability of the laser system. In addition, further pulse shortening through an additional post-compression stage could enable access to the few-cycle regime \cite{Rajhans2023}. In this limit, the laser-plasma interaction would approach the blowout regime, eventually offering enhanced control over the injection process and potentially enabling narrower, tuneable energy distributions at high repetition rates.
Together, these results position industrial high-average-power lasers and multi-pass cell post compression as a promising platform for scalable laser-plasma accelerators capable of delivering the high average particle flux required for next-generation radiation sources.

\begin{acknowledgments}
The authors thank Simon Bohlen and Albrecht Leuschner for their support with the radiation detectors. They additionally thank Soeren Jalas and Manuel Kirchen for the useful conversations. Special thanks goes to the staff of the German Electron Synchrotron DESY, particularly to the Photon Science electronics team and to the Machine Control System team, who provided, mechanical, electrical, computational and administrative support. This work was supported by the Maxwell computational resources at DESY. The authors acknowledge DESY (Hamburg, Germany), a member of the Helmholtz Association HGF, for support and for the provision of experimental facilities.
\end{acknowledgments}

\bigskip\noindent
\section*{Funding}
This work was supported by the German BMFTR, Project 13K2300071 and by the Deutsche Forschungsgemeinschaft (DFG, German Research Foundation), Project 545612524.

\bigskip\noindent
\section*{Declarations}

\subsection*{Author information}
\paragraph*{Authors and affiliations}
\bigskip\noindent

\textbf{Deutsches Elektronen-Synchrotron DESY, Hamburg, Germany}: B. Farace, R.J. Shalloo, T.G. Pak, N. Khodakovskiy, E. Escoto, S. Rajhans, A. Sch\"{o}nberg, I. Hartl, J. Osterhoff, C.M. Heyl, A.R. Maier, K. P\~{o}der and W.P. Leemans.

\textbf{Department of Physics, University of Hamburg, Hamburg, Germany}: W.P. Leemans.

\textbf{Helmholtz-Institute Jena, Jena, Germany}: C.M. Heyl.

\textbf{GSI Helmholtzzentrum für Schwerionenforschung GmbH, Darmstadt, Germany}: C.M. Heyl.

\paragraph*{Contributions}
\bigskip\noindent

B.F., T.G.P and N.K. performed the experiment, with support from K.P.. N.K., E.E, S.R and A.S. designed, built, developed and maintained the laser post-compression system, with support from C.M.H.. B.F., R.J.S, T.G.P, N.K., E.E and S.R designed and built the laser beamline, the vacuum apparatus, the gas delivery system, the electron acceleration chamber and its related diagnostics. B.F. analysed the data and produced the figures, with support from N.K.. B.F. designed the plasma source, performed the particle in cell simulations and wrote the manuscript. W.P.L, C.M.H. and K.P. supervised the personnel and the project, with support from J.O., I.H. and A.M.. The experiment was originally conceived by W.P.L.. All the authors discussed the results in the paper and contributed to the manuscript with relevant comments.

\paragraph*{Corresponding author}
\bigskip\noindent

Correspondence to B. Farace

\subsection*{Ethics declarations}
The authors declare no competing interests.

\subsection*{Code and data availability}
Data and codes underlying the results presented in this paper are not publicly available at this time but may be obtained from the authors upon reasonable request.

\bigskip\noindent

\section*{Methods}\label{Methods}

\subsection*{FBPIC simulations}\label{Methods_0.1}

The laser plasma interaction is simulated with the cylindrical quasi-3D PIC code FBPIC \cite{Lehe2016}, in a Lorentz boosted frame with $\gamma$ = 2, using a resolution of $\Delta$Z = 0.05 $\mu$m and $\Delta$R = 0.1 $\mu$m, with 4 azimuthal modes and with 48 macroparticles per cell.
The laser pulse used has been reconstructed from experimental measurements with LASY \cite{Thevenet2024} using a cylindrical r-t grid with resolutions of $\Delta$t = 0.4 fs and $\Delta$R = 0.08 $\mu$m.

\subsection*{Calibration of the scintillation screen}\label{Methods_0.2}

The electron beam is detected using a DRZ-High scintillation screen based on terbium-doped gadolinium oxysulfide (Gd$_2$O$_2$S:Tb), providing high-efficiency visible emission centred at 545 nm. The screen is mounted in a 2 inches lens mount (Thorlabs) and positioned 16.5 cm downstream of the plasma source, along the laser propagation axis. A thin aluminium foil ($\sim$ \SI{30}{\micro\metre}) is placed before the screen to suppress residual laser light, while having a negligible impact on the electron beam. The absolute light yield of the phosphor screen has been calibrated at the ELBE accelerator in Dresden \cite{Schwinkendorf2019}. To transfer this calibration to the present setup, the collection efficiency of the imaging system is determined via cross-calibration, using a stable tritium light source. The source provides a constant and well-characterised emission, allowing the response of the imaging system to be quantified independently of its optical efficiency. To this end, a tritium source (previously cross-calibrated following the procedure detailed in Appendix A of \cite{Bohlen2021}) is placed at the position of the phosphor screen and used for the current calibration. Its emission is recorded for different exposure times, while keeping the camera gain fixed to the value used during the experiment and using an 8-bit pixel depth. The resulting calibration is therefore directly applicable to the experimental operating conditions, yielding a conversion factor of 5.13 $\times$ 10$^5$ $\pm$ 0.5 $\times$ 10$^5$ counts/pC.

\subsection*{Electron spectrum reconstruction}\label{Methods_0.3}

The electron energy spectrum is reconstructed using a DRZ-High scintillation screen equipped with 12 tungsten filters of varying thickness (25-300 $\mu$m) and exploiting their differential attenuation of the electron signal. The transmitted signal is measured across filtered and unfiltered regions of the screen, and corrected for spatial non-uniformities in the beam profile. Guided by particle-in-cell simulations, the electron spectrum is modelled using an exponentially modified Gaussian distribution and the expected signal for each filter is calculated using energy-dependent transmission functions obtained from GEANT4 simulations \cite{Agostinelli2003}. The spectral parameters are determined by minimising the deviation between measured and modelled transmission values. The reconstructed spectrum is then converted into an absolute charge density by accounting for the energy-dependent response of the scintillator. Uncertainties are estimated through a bootstrap resampling procedure, applied to the measured transmission data.

\subsection*{Plasma density}\label{Methods_0.4}

Plasma density characterisation is performed using Mach-Zehnder transverse interferometry. The electron density profile is measured above the nozzle for various nozzle relative positions and backing pressures. At the highest backing pressures, and when the laser propagation axis is positioned close to the nozzle outlet, the interferometric fringe shift exceeded the retrieval limit, preventing a reliable density reconstruction. In these cases, the density profile is instead estimated using ANSYS-Fluent computational fluid dynamics simulations, benchmarked against the interferometric measurements performed at lower backing pressures, where reconstruction is possible. The resulting density profiles were subsequently used as input for the FBPIC simulations.

\section*{Supplementary Information}

\subsection*{Electron spectrum reconstruction}\label{Supp_0.1}

\begin{figure}
    \centering
    \includegraphics[width=1\linewidth]{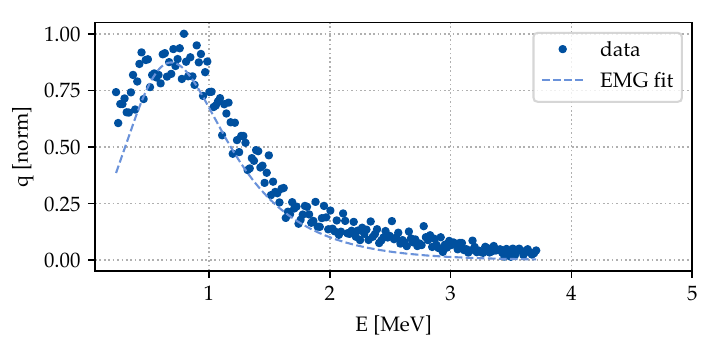}
    \caption{Typical simulated electron spectrum and its corresponding EMG fit.}
    \label{fig:emg_fit}
\end{figure}

The electron energy spectrum is reconstructed using a DRZ-High scintillation screen. To this end, the screen is equipped with 12 tungsten foils (99.95\% purity, HMW Hauner GmbH \& Co. KG), each of $\sim$ 5$\times$5 mm$^2$, with different thickness, ranging from \SI{25}{\micro\metre} to \SI{300}{\micro\metre}. Assuming approximate spatial uniformity of the electron distribution across the screen and a known spectral shape, the differential attenuation provided by the foils can be exploited to reconstruct the electron spectrum.
In this context, Fig.~\ref{fig:emg_fit} shows a typical electron spectrum obtained from FBPIC simulations. The distribution exhibits a low-energy peak followed by an exponential decay extending to a few MeV, which can be accurately described by a three-parameter model. Therefore, guided by the simulation results, the electron spectrum is modelled using an exponentially modified Gaussian (EMG) distribution:

\begin{equation} \label{eq:emg_function}
\begin{split}
EMG(E, \mu, \sigma, \lambda) = &\frac{\lambda}{2} e^{(\lambda (\mu + \lambda \sigma^2/2 - E))} \\
&\text{erfc} \left( \frac{\mu + \lambda\sigma^2 - E}{\sqrt2 \sigma} \right),
\end{split}
\end{equation}

where $\mu$ and $\sigma$ are respectively the mean and the width of the underlying gaussian distribution, $\lambda$ is the exponential decay rate and erfc is the complementary error function.

\begin{figure}
    \centering
    \includegraphics[width=1\linewidth]{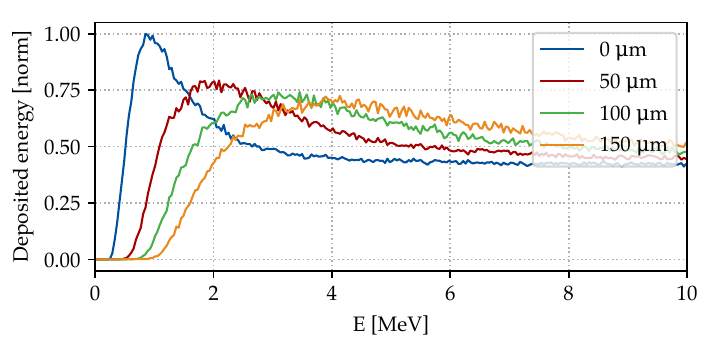}
    \caption{Energy deposited on the scintillator layer of the DRZ-High phosphor screen as a function of the incident electron energy for different tungsten filter thicknesses. The curves shown are the average of 1000000 events simulated in GEANT4.}
    \label{fig:geant4}
\end{figure}

To compute the energy deposited in the scintillating layer of the DRZ-High screen, a series of simulation is performed with the GEANT4 simulation tool \cite{Agostinelli2003}. Here, the electron beam is modelled as a point source with divergence matching the experimentally measured value and propagated through a thin aluminium layer and a set of tungsten filters with varying thicknesses. These simulations serve to extract the energy-dependent transmission of the tungsten filters, which forms the basis of the spectral reconstruction. Some of the resulting transmission curves are shown in Fig.~\ref{fig:geant4} .

\begin{figure}
    \centering
    \includegraphics[width=1\linewidth]{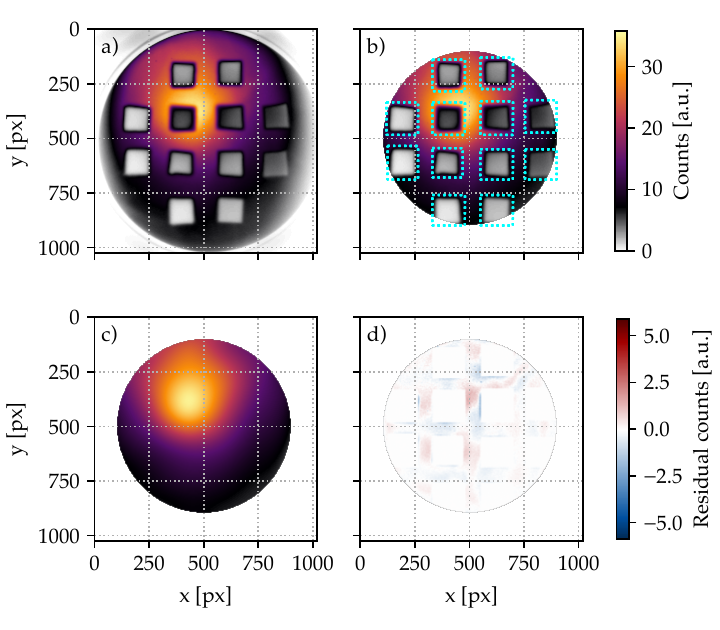}
    \caption{Typical electron screen profile image, averaged over 200 laser shots. Panel a-b: original image and masked image, forming the basis of the analysis. The regions marked in light blue are masked out for the reconstruction of the beam profile. Panels c-d: reconstructed electron beam profile and corresponding residuals.}
    \label{fig:eprofile_screen}
\end{figure}

Fig.~\ref{fig:eprofile_screen}a shows a typical profile screen image averaged over 200 laser shots. As a first step, a circular mask is applied to isolate the emission originating from the scintillator screen Fig.~\ref{fig:eprofile_screen}b.
Next, the electron distribution is reconstructed in the absence of the tungsten filters. Specifically, the regions shadowed by the tungsten attenuators are masked and treated as missing data. The underlying electron fluence is expected to vary smoothly across the screen, as no physical mechanism is expected to produce sharp spatial discontinuities on the scale of the filters features. Therefore, the masked regions are reconstructed using a 2D spline interpolation, constrained by the surrounding data. In contrast to parametric fitting approaches, this method does not impose symmetry or a predefined functional form, thereby minimising reconstruction biases. The procedure is found to be robust with respect to the spline smoothness parameter, yielding negligible variations in both the reconstructed fluence and the derived beam parameters. The resulting reconstructed distribution and it corresponding residuals with respect to the original image are shown in Fig.~\ref{fig:eprofile_screen}c,d.
From here onwards, for simplicity, the original and reconstructed images are denoted by "o" and "r" respectively.

\begin{figure}
    \centering
    \includegraphics[width=1\linewidth]{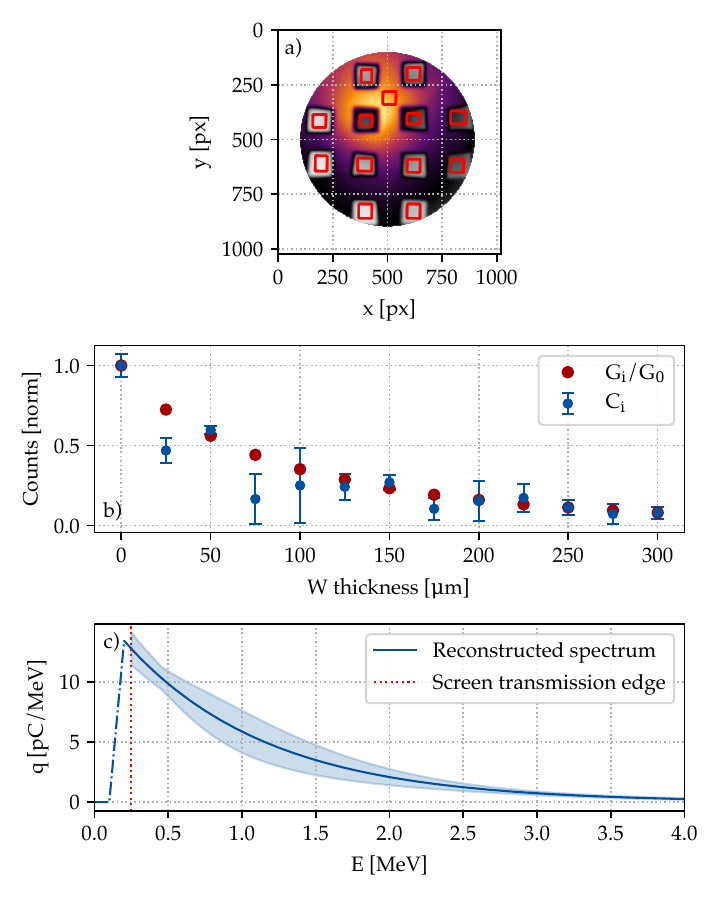}
    \caption{Spectrum retrieval procedure. Panel a: selected areas to extract the mean transmitted counts for each tungsten filter. The area corresponding to W$t$=0 is chosen in a region of strong signal. Panel b: optimum modelled transmission for each tungsten filter (red), compared with the experimental measurement (blue), as described in the main text. Panel c: reconstructed average charge density spectrum.}
    \label{fig:spectral_reconstruction}
\end{figure}

As shown in Fig.~\ref{fig:spectral_reconstruction}a, the average signal within each tungsten-filtered region is extracted over a squared area. For the reference point without a tungsten filter, a region of comparable size is selected on the screen where the signal is high and no attenuation is present. The same regions are evaluated on both the original and reconstructed images, yielding two sets of average counts, $I^o_i$ and $I^r_i$ respectively, where i is thickness of the corresponding tungsten filter.

Owing to the non-uniform spatial distribution of the electron beam on the screen, the effective transmission of each filtered region depends on the local fluence, and is therefore sensitive to beam pointing. To account for this effect, the measured counts $I^o_i$ are rescaled by a correction factor $k_i = I_{peak} / I^r_i$, where $I_{peak}$ is the peak signal on the original image. The corrected counts are then normalised to the reference signal obtained in the unfiltered region. This procedure yields the set of normalised transmission values: 

\begin{equation} \label{eq:normalised_counts}
C_i=\frac{I^o_i k_i}{I^o_0 k_0},
\end{equation}

where $i=0$ corresponds to the region without any tungsten filter.

The reconstruction procedure then consists in determining the set of parameters that best reproduces the measured transmission values. In particular, the EMG function defined in Eq.~\ref{eq:emg_function} is used to model the electron energy spectrum. For a given set of parameters $(\mu,\lambda,\sigma)$, the expected signal for each tungsten filter is computed by applying the GEANT4-derived transmission.
The optimal parameters are obtained by minimising the weighted residuals between the model and the measured data using a least-squares approach. Specifically, the residuals are defined as $G_i/G_0 - C_i$ and are weighted over the experimental standard deviation associated with each $C_i$. Here, the modelled signal is given by 

\begin{equation} \label{eq:modelled_counts}
G_i=\int_E EMG(E,\mu,\lambda,\sigma) t_i(E) dE, 
\end{equation}

where $t_i$ denotes the transmission through a tungsten filter of thickness i, as obtained from the GEANT4 simulations. The results of this optimisation procedure are shown in Fig.~\ref{fig:spectral_reconstruction}b, where the normalised modelled signal $G_i/G_0$ is compared, for each tungsten thickness, to its corresponding measured signal $C_i$.  

The final step consists in converting the normalised spectrum into an absolute charge density (pC/MeV), which requires an accurate estimate of the total accelerated charge. Here, a direct conversion from camera counts using the previous calibration is not sufficient, as the tritium-based cross-calibration is referenced to electron energies of the order of 40 MeV \cite{Schwinkendorf2019}. At lower energies, the response of the DRZ-High scintillator becomes energy dependent, approaching a constant value only above $\sim$~5 MeV. This effect is accounted for by combining the reconstructed spectral distribution with the energy-dependent scintillator response, enabling a consistent determination of the total charge and, hence, of the absolute charge density (Fig.~\ref{fig:spectral_reconstruction}c).

The uncertainty on the reconstructed electron spectrum is estimated using a bootstrap resampling method \cite{Efron1994}. While parameter uncertainties could in principle be derived from the covariance matrix of the least-squares fit, this approach is not reliable in the present case due to the strong nonlinearity of the forward model and the correlation between fit parameters. As a result, the inverse problem is only weakly constrained, and small variations in the input data can lead to large variations in the fitted parameters, yielding unrealistically large covariance estimates.
To obtain a more robust estimate of the uncertainty, a bootstrap procedure is employed. A thousand synthetic datasets are generated by resampling the measured transmission values $C_i$ assuming Gaussian noise consistent with their experimental uncertainties, and the full reconstruction procedure is repeated for each realisation. The resulting ensemble of spectra is then used to extract confidence intervals on the reconstructed electron distribution.

\subsection*{Electrons injection}\label{Supp_0.3}

\begin{figure}[ht]
    \centering
    \includegraphics[width=1\linewidth]{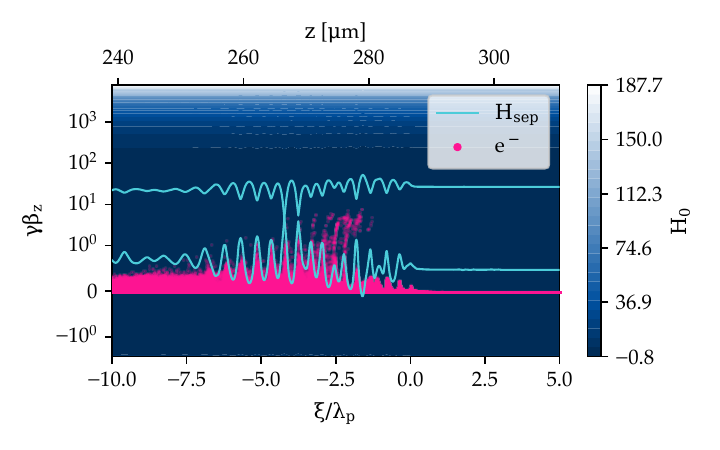}
    \caption{Hamiltonian representation of the plasma wakefield at the injection instant in the normalised co-moving frame. The separatrix trajectory H$_{sep}$ and the plasma electrons are shown respectively in light blue and pink. The injected electrons are the one falling in the separatrix contour. The longitudinal momentum axis $\gamma \beta_z$ is plotted in symmetric logarithmic scale.}
    \label{fig:fbpic_injection}
\end{figure}

The Hamiltonian phase-space distribution shown in Fig~\ref{fig:fbpic_injection} is instrumental to visualise the electron injection instant. The representation is given in the normalised co-moving frame, where $\xi=ct - z$ and $\gamma\beta_z = p_z/m_ec$ is the normalised longitudinal electron momentum. In this framework, electron injection into the wakefield occurs when the longitudinal momentum exceeds that of the separatrix orbit: 

\begin{equation} \label{eq:separatrix}
H_{sep} =  \sqrt{1+\gamma_g^2\beta_g^2+a^2(\xi_{min})}-\phi(\xi_{min})-\gamma_g\beta_g^2,
\end{equation}

where $\xi_{min}$ is the position of the minimum of the wakefield normalised scalar potential $\phi$ and $\gamma_g$ is the Lorentz factor associated with the normalised laser group velocity $\beta_g$. As evident from the phase-space structure, the resonant excitation of the wakefield leads to injection across multiple plasma buckets, ultimately resulting in a broad energy distribution.


%

\end{document}